\documentclass[prl,showpacs,twocolumn,aps,superscriptaddress,a4paper]{revtex4}
\usepackage{dcolumn}
\usepackage{amsmath}
\usepackage{graphicx}
\usepackage{latexsym}  
\usepackage{amsfonts}  
\usepackage{amssymb}  
\DeclareGraphicsExtensions{.eps,.pdf,.gif,.jpg}  
  
\newcommand{\be}{\begin{equation}}  
\newcommand{\ee}{\end{equation}}
\newcommand{\beq}{\begin{eqnarray}}  
\newcommand{\eeq}{\end{eqnarray}}

\begin{document}  
      
\def\bbe{\mbox{\boldmath $e$}}  
\def\bbf{\mbox{\boldmath $f$}}      
\def\bg{\mbox{\boldmath $g$}}  
\def\bh{\mbox{\boldmath $h$}}  
\def\bj{\mbox{\boldmath $j$}}  
\def\bq{\mbox{\boldmath $q$}}  
\def\bp{\mbox{\boldmath $p$}}  
\def\br{\mbox{\boldmath $r$}}      
  
\def\bone{\mbox{\boldmath $1$}}      
  
\def\dr{{\rm d}}  
  
\def\tb{\bar{t}}  
\def\zb{\bar{z}}  
  
\def\tgb{\bar{\tau}}

\def\bC{\mbox{\boldmath $C$}}  
\def\bG{\mbox{\boldmath $G$}}  
\def\bH{\mbox{\boldmath $H$}}  
\def\bK{\mbox{\boldmath $K$}}  
\def\bM{\mbox{\boldmath $M$}}  
\def\bN{\mbox{\boldmath $N$}}  
\def\bO{\mbox{\boldmath $O$}}  
\def\bQ{\mbox{\boldmath $Q$}}  
\def\bR{\mbox{\boldmath $R$}}  
\def\bS{\mbox{\boldmath $S$}}  
\def\bT{\mbox{\boldmath $T$}}  
\def\bU{\mbox{\boldmath $U$}}  
\def\bV{\mbox{\boldmath $V$}}  
\def\bZ{\mbox{\boldmath $Z$}}  
  
\def\bcalS{\mbox{\boldmath $\mathcal{S}$}}  
\def\bcalG{\mbox{\boldmath $\mathcal{G}$}}  
\def\bcalE{\mbox{\boldmath $\mathcal{E}$}}  
  
\def\bgG{\mbox{\boldmath $\Gamma$}}  
\def\bgL{\mbox{\boldmath $\Lambda$}}  
\def\bgS{\mbox{\boldmath $\Sigma$}}  
  
\def\bgr{\mbox{\boldmath $\rho$}}  
  
\def\a{\alpha}  
\def\b{\beta}  
\def\g{\gamma}  
\def\G{\Gamma}  
\def\d{\delta}  
\def\D{\Delta}  
\def\e{\epsilon}  
\def\ve{\varepsilon}  
\def\z{\zeta}  
\def\h{\eta}  
\def\th{\theta}  
\def\k{\kappa}  
\def\l{\lambda}  
\def\L{\Lambda}  
\def\m{\mu}  
\def\n{\nu}  
\def\x{\xi}  
\def\X{\Xi}  
\def\p{\pi}  
\def\P{\Pi}  
\def\r{\rho}  
\def\s{\sigma}  
\def\S{\Sigma}  
\def\t{\tau}  
\def\f{\phi}  
\def\vf{\varphi}  
\def\F{\Phi}  
\def\c{\chi}  
\def\w{\omega}  
\def\W{\Omega}  
\def\Q{\Psi}  
\def\q{\psi}  
  
\def\ua{\uparrow}  
\def\da{\downarrow}  
\def\de{\partial}  
\def\inf{\infty}  
\def\ra{\rightarrow}  
\def\bra{\langle}  
\def\ket{\rangle}  
\def\grad{\mbox{\boldmath $\nabla$}}  
\def\Tr{{\rm Tr}}  
\def\Re{{\rm Re}}  
\def\Im{{\rm Im}}

\title{Conserving approximations in
time-dependent quantum transport: Initial correlations and memory 
effects}

\author{Petri My\"oh\"anen}
\affiliation{Department of Physics, Nanoscience Center, FIN 40014, University of Jyv\"askyl\"a,
Jyv\"askyl\"a, Finland}
\author{Adrian Stan}
\affiliation{Department of Physics, Nanoscience Center, FIN 40014, University of Jyv\"askyl\"a,
Jyv\"askyl\"a, Finland}
\author{Gianluca Stefanucci}
\affiliation{Dipartimento di Fisica, Universit\`a di Roma Tor Vergata, Via della Ricerca Scientifica 1, I-00133 Rome, Italy}
\affiliation{European Theoretical Spectroscopy Facility (ETSF)}
\author{Robert van Leeuwen}
\affiliation{Department of Physics, Nanoscience Center, FIN 40014, University of Jyv\"askyl\"a,
Jyv\"askyl\"a, Finland}
\affiliation{European Theoretical Spectroscopy Facility (ETSF)}

\date{\today}  

\begin{abstract}
We study time-dependent quantum transport in a correlated model system
by means of time-propagation of the Kadanoff-Baym equations for the nonequilibrium many-body Green function.
We consider an initially contacted equilibrium system of a correlated central region coupled 
to tight-binding leads. Subsequently a time-dependent bias is switched on after which we follow in detail
the time-evolution of the system. Important features of the Kadanoff-Baym approach are 1) 
the possibility of studying the ultrafast dynamics of transients and 
other time-dependent regimes and 2) the inclusion of exchange 
and correlation effects in a conserving approximation scheme.
We find that initial correlation and memory terms due to many-body interactions have a large effect on the transient currents. Furthermore the value of the steady state current is found to be strongly dependent on the 
approximation used to treat the electronic interactions.
\end{abstract}
  
  
\maketitle  

The ultimate goal of molecular 
electronics~\cite{book} in solid state circuitry is
to miniaturize the size and maximize the speed of integrated devices.
Advances in this field crucially depend on the accumulated experimental 
and theoretical knowledge. For the latter to progress it is essential 
to develop quantum mechanical approaches that are able to deal 
with {\em open} and {\em interacting} systems in an {\em out of 
steady-state regime}.
Desirable features of such approaches are therefore 1) 
the possibility to study the ultrafast dynamics of transients and 
other time-dependent (TD) regimes and 2) the inclusion of exchange 
and correlation effects in a {\em conserving approximation scheme}.
Feature 1) was incorporated in some recently proposed 
one-particle frameworks and was exploited to address several issues 
in TD quantum transport (QT)~\cite{ksarg.2005,zmjg.2005,hhlkch.2006,mgm.2007}. 
These frameworks can, in principle, be combined 
with TD density functional theory~\cite{rg.1984,sa.2004,sa2.2004,dvt.2004}, 
thus providing a route to include 
Coulomb interactions (possibly in a conserving way~\cite{vbdvls.2005}). 
Feature 2) is an essential 
requirement as realistic time evolutions must preserve 
basic conservation laws as, for instance, the continuity 
equation. Conserving approximations~\cite{b.1962} like, e.g. self-consistent 
Hartree-Fock (HF), second Born (2B) or GW, have recently been employed 
in the context of QT but the implementations have been, sofar, 
restricted to steady-state regimes~\cite{tr.2007,dfmo.2007,wshm.2008,tr.2008,t.2008}.\\
In this Letter we propose an alternative approach to TD-QT that 
encompasses both feature 1) and 2). It is 
based on the real-time propagation of the {\em embedded} Kadanoff-Baym 
(KB) equations~\cite{book2,dvl.2007,d.1984,kb.2000} which are equations of motion
for the nonequilibrium Green function from which basic properties
of the system can be calculated.
We consider a set $\{\a\}$ of {\em noninteracting} electronic reservoirs 
connected via a tunneling Hamiltonian to 
an {\em interacting} many-body quantum system $C$. The Green function $G(z,z')$ 
(we suppress basis indices)
projected on $C$ obeys the equation of motion~\cite{dvl.2007,d.1984}
\be
[i\de_{z}-h(z)]G(z,z')=\d(z,z')+\int_{c}d\bar{z}\
\S(z,\bar{z})G(\bar{z},z')
\label{eom}
\ee
where $z$ and $z'$ are time-coordinates on the 
Keldysh contour $c$~\cite{d.1984}. We consider systems initially 
(times $t < 0$) contacted and in equilibrium at inverse temperature $\beta$
and chemical potential $\mu$. The corresponding contour is
described in Refs. \onlinecite{sa.2004} and \onlinecite{dvl.2007}.
In Eq.(\ref{eom}),
$h(z)$ is the
one-body Hamiltonian of the interacting system $C$
and $\S$ is the time-nonlocal self-energy. The latter
describes the effects of many-body interactions and
embedding of the system and is
the sum of a many-body self-energy $\S_{\rm MB}[G]$ and 
an embedding self-energy $\S_{\rm emb}$. The former is a functional of 
the projected Green function $G$ only and can be expressed
in terms of Feynman diagrams while the latter is a sum $\S_{\rm emb}=\sum_{\alpha} \S_{\rm emb, \alpha}$ 
where
\be
\S_{\rm emb,\alpha}(z,z')= t_{C\a}(z)g_{\a}(z,z')t_{\a C}(z').
\label{emb}
\ee
In Eq.(\ref{emb}) $g_{\a}$ is the Green function of the 
uncontacted lead $\a$ and matrices $t_{C\a}$ and $t_{\a C}$
describe the couplings of system $C$ to the leads.
The TD equations obtained from Eq.(\ref{eom}) by restricting time-arguments
to different parts of the Keldysh contour are known as the
Kadanoff-Baym equations~\cite{book2,dvl.2007,d.1984,kb.2000} and are the main
tools of this work. 
As the system is driven out of equilibrium by a TD bias voltage, the current flowing into lead $\alpha$
is obtained by taking the time derivative of the total number of particles in $\alpha$~\cite{mw.1992} and reads
\begin{eqnarray}
I_{\a}(t) &=&-2 \Re\,\Tr_{C} [ G^< \cdot \Sigma_{\a,\rm emb}^{\rm A}
+ G^{\rm R}\cdot \Sigma_{\a,\rm emb}^{<}] (t,t) \nonumber \\
&& - 2 \Re\, \Tr_{C} [ G^{\rceil} \star
\Sigma_{\a,\rm emb}^{\lceil} ](t,t)
\label{current}
\end{eqnarray}
where the trace is taken over the central region indices
and the products $\cdot$ and $\star$ denote integrations over
the real and imaginary tracks of the contour (see Ref.\onlinecite{sa.2004}
for details).
The objects superindexed with $\lessgtr, \rceil \lceil$
correspond to time arguments
on different parts of the Keldysh contour~\cite{sa.2004,dvl.2007} and
R/A denote the retarded and advanced components.
The last term in Eq.(\ref{current}) arises from integration along
the imaginary branch $(0,-i\beta)$ of the Keldysh contour~\cite{sa.2004,dvl.2007}
and explicitly accounts for the effects of
initial correlations and initial-state dependence. If one assumes that
both dependencies are washed out in the long-time limit ($t \rightarrow \infty$)
then the last term in Eq.(\ref{current}) vanishes and the Meir-Wingreen 
formula~\cite{mw.1992}
is recovered. The KB approach provides a natural tool to investigate
the validity of this assumption which has remained
unexplored sofar, and that we partly address below.\\
Using the KB equations we first solve
the embedded and correlated equilibrium problem and then propagate the 
system in time after applying a time-dependent bias (cf. Fig.1 of Ref.\onlinecite{dvl.2007}). 
For this we extend the
implementation of Ref.\onlinecite{dvl.2007} to open systems, i.e., by replacing 
$\S_{\rm MB}[G]$ with $\S_{\rm MB}[G]+\S_{\rm emb}$.
Different TD perturbations allow us to address 
several open issues in correlated TD-QT. 1) We can set the tunneling 
Hamiltonian to zero at the initial time or not
(which corresponds to $C$ initially contacted or uncontacted) 
and study the effects of the initial conditions 
(partition-free~\cite{c.1980} vs. contacting approach~\cite{ccnsj.1971}) 
on the TD density and total current.
2) Many-body interactions can be included in the initial 
solution of the equilibrium problem or switched on at later times.
In this way effects of initial correlations~\cite{d.1984}  
on transient and steady-state properties can be 
highlighted. 3) The dependence on the history of the applied bias 
can be investigated 
for different approximate $\S_{\rm MB}[G]$'s. 
Due to the nonlinearity of the problem and possibly to the 
nonlocality in time of the many-body self-energy nontrivial memory effects 
may occur (bistability, hysteresis phenomena, etc.). 
4) We can study ac biases, pulses or other kind of TD biases as well as TD gate 
voltages and TD contacts in correlated QT. We
stress that for a given approximate $\S_{\rm MB}[G]$ all kinds of TD 
perturbations within the KB 
approach require the same computational effort.\\
To study several of the issues mentioned in points 1)-4) above we consider an interacting device coupled to noninteracting one-dimensional leads. The full system is described by:
\begin{eqnarray}
\hat{H}(t) &=& \sum_{ij,\sigma \alpha} [t_{ij}^{\alpha} +\delta_{ij}U_{\alpha}(t)] \hat{c}_{i\sigma\alpha}^{\dagger}\hat{c}_{j\sigma \alpha} + \nonumber \\
&&\sum_{ij,\sigma} t_{ij}\hat{d}_{i\sigma}^{\dagger}\hat{d}_{j\sigma} + \frac{1}{2}\sum_{ij,\sigma\sigma'} v_{ij}\hat{d}_{i\sigma}^{\dagger}\hat{d}_{j\sigma'}^{\dagger}\hat{d}_{j\sigma'}\hat{d}_{i\sigma} + \nonumber \\
&&\sum_{ij,\sigma \alpha} V_{i,j\alpha} [ \hat{d}_{i\sigma}^{\dagger}\hat{c}_{j\sigma \alpha} + 
\hat{c}_{j\sigma \alpha}^{\dagger}\hat{d}_{i\sigma} ]
\label{ham}
\end{eqnarray}
where $i,j$ are site indices, $\sigma, \sigma'$ spin indices and
where $\hat{c}^{\dagger},\hat{c}$ and $\hat{d}^{\dagger},\hat{d}$ are the creation and annihilation
operators for leads and central region respectively. 
The first term in Eq.(\ref{ham}) describes the leads with a TD bias $U_{\alpha}(t)$ while the second and third term 
describe the one-body and many-body interactions of the central region. 
Finally the last term describes the coupling between the leads and the central region.
Since we consider semi-infinite leads, the energy band of the leads is continuous with a finite band width
and we therefore do not employ the often used wide band limit approximation.
We stress that the KB approach is not limited
to 1-D leads as the leads enter only via the embedding self-energy.\\
\begin{figure}[t]
\flushleft
\includegraphics[width=0.45\textwidth]{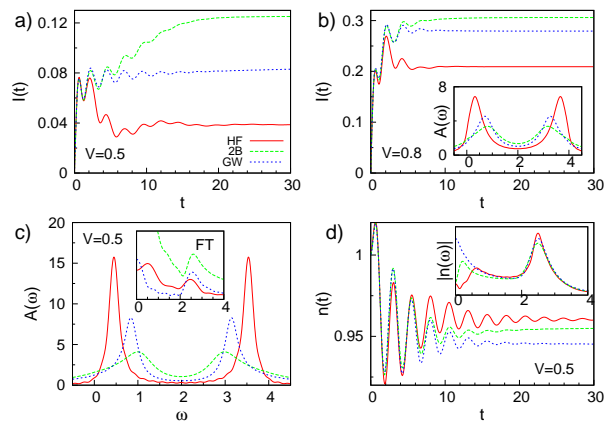}
\caption[]{(color online) 
Transient currents for the HF (red),2B (green) and GW (blue) approximations $V=0.5$ (a) and $V=0.8$ (b). 
The spectral functions are displayed in c) ($V=0.5$) and in the inset of b) ($V=0.8$).
The inset in c) depicts the modulus of the Fourier transform of the transients. 
d) density occupation number $n(t)$ on site 1. The inset
shows the modulus $|n(\omega)|$ of its Fourier transform}
\label{fig1}
\end{figure}
We consider a system with two central sites coupled to leads located on the left and right
of the central region (i.e. $\alpha=L,R$).
We use the parameters $t_{11}=t_{22}=0$ and $t_{12} = t_{21}=-1$.
For the many-body interactions we take $v_{11}=v_{22}=2$ and $v_{12}=v_{21}=1$. 
The chemical potential $\mu$ for the whole initial equilibrium system is set at the middle of the HF gap and is determined by a Hartree-Fock calculation on the uncontacted but correlated central region which yields the value $\mu=2$ with the parameters described above.
For the semi-infinite leads
we use $t^{L/R}_{ii}=\mu$ and $t^{L/R}_{ij}=-1.5$ if $i$ and $j$ are neighboring sites and zero otherwise. 
The leads are coupled to the central region by coupling elements $V_{1,jL}=V_{2,jR}=V$
when $j$ is the first site on the left or right lead and zero otherwise 
(we use $V=0.5$ and $V=0.8$). We further take $\beta=90$ (zero temperature limit).
All quantities are expressed in atomic units.
For $\S_{\rm MB}$ we employ the HF, 2B and GW
approximations which have also been used in earlier transport studies~\cite{tr.2008,t.2008}.
The GW approximation includes the dynamical screening of the electron interaction
whereas 2B includes all Feynman diagrams to second order in the bare interaction.
In the small system that we study here the second order exchange diagram (incorporated in 2B
but not in GW) is not negligible and therefore the 2B approximation gives probably the most accurate 
description of electronic correlations.\\
\textit{Correlations in transients}. In Fig.\ref{fig1} we show the transient currents
flowing into the right lead and the spectral functions for the HF, 2B and GW approximations. 
The system is driven by a symmetrically 
switched bias $U_L(t)=-U_R(t)=U \theta(t)$ with $U =1.0$, where $\theta(t)$ is a Heaviside function, i.e. we consider
a sudden switch on of the bias at $t=0$. Results are displayed for weak 
($V=0.5$) and strong coupling ($V=0.8$) of the central region to the leads. 
We first consider the spectral functions which are defined as
$A(T,\omega) = -\textnormal{Tr Im} \int dt\, e^{i\omega t}[G^> - G^<](T+\frac{t}{2},T-\frac{t}{2})$
where $T=(t_1+t_2)/2$ and $t = t_1-t_2$.
We find that after the steady state has been reached the spectral functions do not depend on $T$ anymore.
In equilibrium the spectral peaks are at $\epsilon_{1,2}^0=0.5,3.5$ for HF, 2B and GW.
For the biased system (Fig.1c and inset of Fig.1b) the
electron correlations beyond HF lead to a narrowing of the gap between 
the spectral peaks and a broadening of the spectral function.
This bias dependent gap closing mechanism was recently identified by Thygesen \cite{t.2008}.
The positions $\epsilon_{1,2}$ of the spectral peaks strongly affect the
final steady state currents as they are largest when both spectral peaks enter
the bias window, i.e. for biases such that $\mu \pm U  \approx \epsilon_{1,2}$.
This condition is much better satisfied for
2B and GW than for HF and explains the higher values in 2B (the highest) and GW (top panels of Fig.\ref{fig1}). 
Let us now focus on the temporal structure of the transients.
The transient currents show an
oscillation that becomes more pronounced when we weaken the coupling from
$V=0.8$ to $V=0.5$.
The modulus of the Fourier transform of the current (minus its steady state value) is
displayed in the inset of Fig.1c. There is a frequency peak
at 2.5 in all many-body approximations and for HF also one at 
0.5 (for 2B and GW there is a broad peak around zero).
These frequencies cannot be directly related to
the spectral functions of Fig.\ref{fig1} as those correspond to the steady state limit
when the transients have settled. The frequencies instead correspond
to transitions between the spectral peaks $\epsilon_{1,2}^0$ of the initial equilibrium system
 (which for HF,2B and GW have similar values of $\epsilon_{1,2}^0-\mu \approx \pm 1.5$)
and the incoming/outgoing states at the left/right Fermi energy $\mu \pm U$.
The peak at $0.5$ is not visible for 2B and GW because the corresponding oscillation is damped
faster than in HF.
In the case $V=0.8$ the current oscillations are suppressed and the steady state is obtained earlier compared to the $V=0.5$ case since the electrons can tunnel in and out of the device more easily.
The oscillations have a clear relation to density changes in the central region.
In Fig.1d we display the TD site occupation  $n_1(t) = -iG_{11}^<(t,t)$ of site 1 in the central region (the occupations on site 1 and 2 satisfy $n_1(t) + n_2 (t) \simeq 2$). The sudden switch-on of the bias generates a density oscillation
in the central region which damps on a time-scale comparable to the time to reach the steady
state current. In this limit the system becomes polarized and part the electron density is accumulated to the right side of the device i.e. in the direction in which the current flows. 
The modulus of Fourier transform of $n_1(t)$ (inset) displays peaks at exactly the same 
frequencies as obtained from the wiggles in the transient currents. \\
{\em Conservation of charge.}
Since we use conserving approximations particle number must be conserved in the system.
This is illustrated
in Fig.2a for the 2B approximation: the 
system is driven out of equilibrium by an asymmetric steplike bias $U_L=0.9$, $U_R=-0.4$.
This plot shows the currents $I_{L/R}(t)$ as well as the time derivative of the number of
particles $N(t)$ in the central region: clearly the particle number conservation law $I_L+I_R = -dN/dt$ is obeyed.\\
\textit{Initial state dependence and memory}. 
Initial correlations manifest themselves in two ways in the KB equations.
First, the initial values of the time-dependent Green functions are determined by
the equilibrium Green function at $t=0$ that is obtained by considering
both time arguments on the vertical track of the Keldysh contour.
Second, the KB equations contain terms that describe memory of the initial state during the
time propagation. These terms depend on the self-energies $\Sigma^{\rceil/\lceil}(z,z')$~\cite{sa.2004,dvl.2007}
with mixed real and imaginary time arguments.
We investigate these two memory effects separately by either
setting the self-energy $\Sigma_{\rm MB} (z,z')$ 
to zero for $z$ and $z'$ on the vertical track of the contour
(initial state is noncorrelated) or by 
setting $\Sigma_{\rm emb}^{\rceil/\lceil}$ and/or $\Sigma_{\rm MB}^{\rceil/\lceil}$ to zero
(initial state is
correlated but memory effects due to embedding and/or electron correlations are neglected).\\
We start by setting $\Sigma_{\rm MB}$ to zero when both time-arguments are on
the vertical track of the contour for a situation in which we propagate an unbiased system
from time $t=0$ to a finite time $t_0$ at which time we switch on a sudden symmetric bias, i.e. we use 
$U_L(t)=-U_R (t) = \theta (t-t_0) U$.
Since electron correlations are taken into account in the time propagation but not in the initial state 
there will be a charge redistribution for times $t > 0$.
The result is compared to an initially correlated KB propagation.
We take $t_0=20$ and $U=1.0$.
The noninteracting system at $t=0$ has
spectral peaks at energies $\epsilon_{1,2}^0= \pm 1.0$ (i.e. below the chemical potential at $\mu=2$).
As a result of electron interactions for $t > 0$ we find upward shifts in the spectral peaks
yielding one peak above and one peak below $\mu$.
As a consequence a charge of about 2 electrons is pushed into the leads.
The corresponding current, shown in Fig.2b for the 2B approximation,
is saturated before the bias voltage is switched on at time $t_0$.
\begin{figure}[t]
\flushleft
\includegraphics[width=0.45\textwidth]{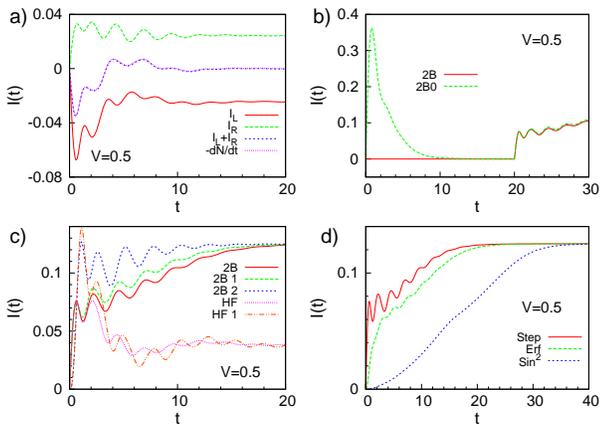}
\caption[]{(color online) a) Transient currents for asymmetric bias (see text).  
b) Transient currents for 2B with and without the initially interacting ground state 
(2B and 2B0 correspondingly).c) Transient currents in HF and 2B approximations with and without 
the memory terms $\Sigma_{\rm emb/ MB}^{\rceil / \lceil}$ (see text).
d) Transient currents for different bias-switchings.
All panels correspond to $V=0.5$ }
\label{fig2}
\end{figure}
For later times $t > t_0$
the transient currents with inclusion and with neglect of initial correlations
are indistinguisable. We therefore conclude
that the initially uncorrelated system has relaxed to a correlated state when
the bias is switched on.\\
To study how initial states are remembered during time-propagation we compare
full solutions of the KB equations to ones in which
we neglect the terms $\Sigma_{\rm emb}^{\rceil/\lceil}$
and/or $\Sigma_{\rm MB}^{\rceil/\lceil}$.
However, at the initial time $t=0$ we still employ the fully correlated embedded
equilibrium Green function.
The results are displayed in Fig.2c
for the HF and 2B approximations. 
We find that neglect of the memory terms $\Sigma^{\rceil/\lceil}$ has a considerable effect on the transient currents. 
In the HF case these terms only contain the
embedding self-energy $\Sigma_{\textrm{emb}}^{\rceil/\lceil}$ (as $\Sigma_{\rm MB}$ of HF is purely local in time) and therefore the term describes memory of the
initial contacting of leads. Neglect of this term leads to the curve labeled HF 1
in Fig.2c. For the 2B case there is also a dependency on the
many-body self-energy $\Sigma_{\rm MB}^{\rceil/\lceil}$. 
We therefore have two curves for 2B, one in which we neglect only $\Sigma_{\textrm{MB}}^{\rceil/\lceil}$ (labeled 2B 1) and one in which we neglect both $\Sigma_{\textrm{emb}}^{\rceil/\lceil}$
and $\Sigma_{\textrm{\rm MB}}^{\rceil/\lceil}$ (labeled 2B 2).
We see that neglect of $\Sigma_{\textrm{emb}}^{\rceil/\lceil}$ has a considerable effect on the transients
while neglect of only $\Sigma_{\rm MB}^{\rceil/\lceil}$ has a smaller but still noticable effect.
We further see that the same steady state current develops as with 
the memory terms included and therefore conclude that
the memory terms eventually die out in the long-time limit.
This is in agreement with the memory loss theorem proven in Refs.\onlinecite{sa.2004} and \onlinecite{sa2.2004}
for the case of Green functions that are sufficiently smooth. We finally note 
that there are situations for which the Green function is not a smooth function
in which case persistent oscillations may appear, see Ref.\onlinecite{s.2007}.\\
\textit{Time dependence of applied bias.}
We finally investigate the dependence of the transient currents on various forms of the
time-dependent bias.
In Fig.2d we show the 2B transient currents driven by different 
TD symmetric biases: $U_L(t)=-U_R(t)$. We take $U_L (t)=U\,\theta(t)$, $U_L(t)=U\,\textnormal{Erf}(\omega_1t)$ 
and $U_L(t)=U \sin^2(\omega_2t)$ for $t\leq \frac{\pi}{2\omega_2}$ and $U_L(t)=U$ for $t>\frac{\pi}{2\omega_2}$
with $U=1.0$, $\omega_1=0.5$ and $\omega_2=0.1$. We observe that the sudden 
switch-on produces rapid oscillations. They are more damped with slower switch on of the bias voltage.
The steady state currents are, however, the same for all three cases.
However, due to nonlinearity of the KB equations existence of bistable solutions and hence different
steady states may be possible for different biases. This will be part of future investigations.\\
We conclude that the KB equations provide a powerful tool to study correlated
quantum transport in real time. The method allows for inclusion of many-body correlations 
while satisfying important conservation laws.
We found that many-body interactions have large effects on steady-state and transient currents.
The temporal features in the transients and density distributions were analyzed in detail and related to level structure displayed in the spectral functions, a study of utmost importance to interpret transport spectroscopy experiments. We further showed that memory terms have large effects on the time-dependent currents.

\end{document}